\documentclass[a4paper]{jpconf}
\usepackage{graphicx}
\usepackage{subfig}
\usepackage{lineno}
\begin{document}
\title{Data Processing at the Pierre Auger Observatory}

\author{Jakub V\'icha$^{1}$, Ji\v{r}\'i Chudoba$^{1}$ for the Pierre Auger Collaboration$^{2}$}

\address{$^{1}$Fyzik\'aln\'i \'ustav AV \v{C}R, Na Slovance 1999/2, Praha 8, 18221, Czech Republic}
\address{$^{2}$Full author list: http://www.auger.org/archive/authors{\_}2014{\_}08.html}
\ead{vicha@fzu.cz}

\begin{abstract}
Cosmic--ray particles with ultra--high energies (above $10^{18}$~eV) are studied through the properties of extensive air showers which they initiate in the atmosphere. The Pierre Auger Observatory detects these showers with unprecedented exposure and precision and the collected data are processed via dedicated software codes. Monte Carlo simulations of extensive air showers are very computationally expensive, especially at the highest energies and calculations are performed on the GRID for this purpose. The processing of measured and simulated data is described, together with a brief list of physics results which have been achieved.
\end{abstract}

\section{Introduction}
\begin{figure}[!ht]
\centering
\includegraphics[width=0.6\textwidth]{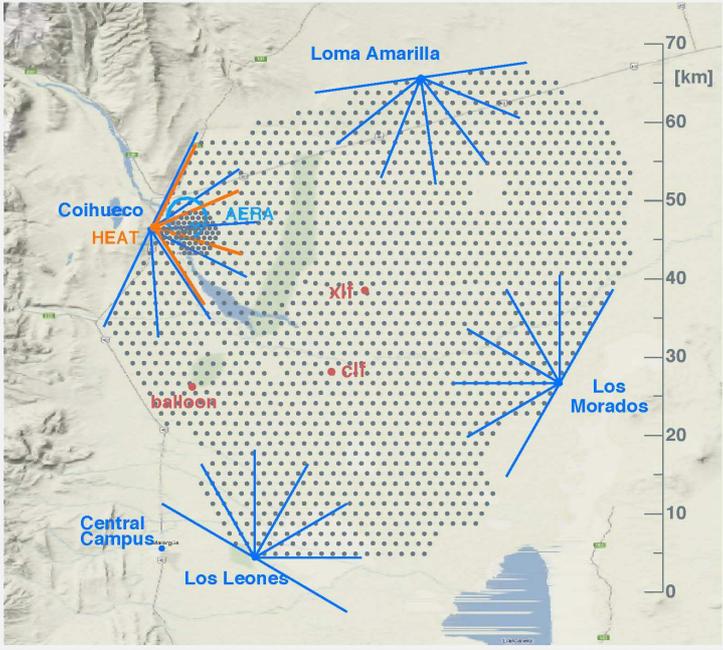}
\caption{The view of the Pierre Auger Observatory deployment. Array of gray points represents positions of surface detector stations. The names of 4 fluorescence detector buildings are shown in blue together with illustrated azimuthal field of views of the individual fluorescence telescopes. Positions of atmospheric monitoring stations are marked in red. Locations of additional radio detection stations (AERA) and FD enhancement (HEAT) are also shown together with the position of the Central Campus.}
\label{Auger_map}
\end{figure}

\begin{figure}[!hb]
\centering
\subfloat{\includegraphics[width=0.6\textwidth]{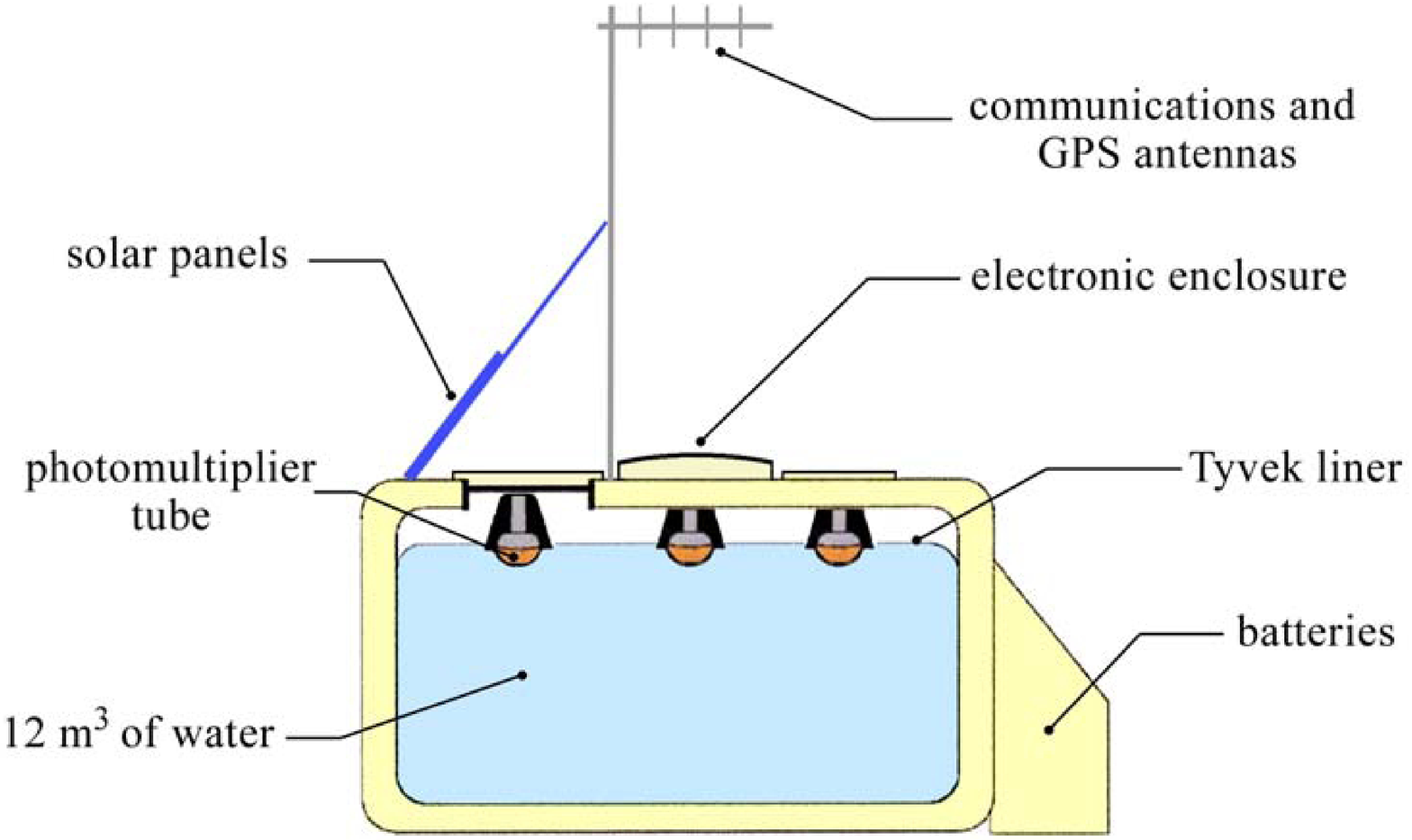}}
\subfloat{\includegraphics[width=0.4\textwidth]{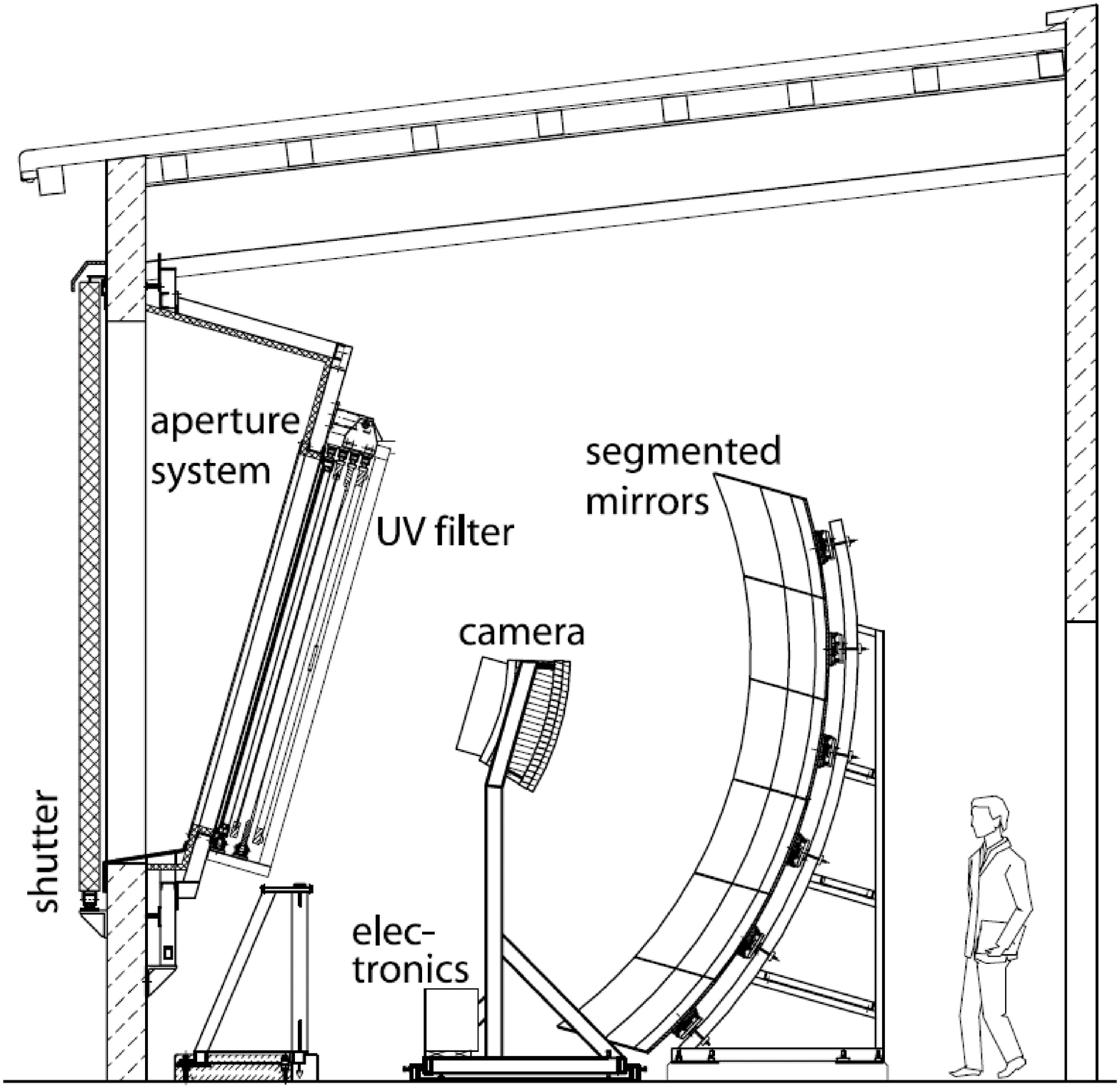}}
\caption{Schematic description of one water--Cherenkov station \cite{Auger} (left panel) and one fluorescence telescope \cite{FdPaper} (right panel).}
\label{Detectors}
\end{figure}

The Pierre Auger Observatory \cite{Auger} is the largest experiment ever built to measure the extensive air showers induced by cosmic rays of energy above $10^{18}$~eV (UHECR). The Observatory (see Fig.~\ref{Auger_map}) is located on the Argentinian Pampa at an altitude of about 1~400~m~a.s.l., near the city of Malarg\"{u}e in the province Mendoza. It predominantly observes the southern sky. On the ground, shower detection employs water--Cherenkov detectors (see left panel of Fig.~\ref{Detectors}). These are combined with the telescopic observation (see right panel of Fig.~\ref{Detectors}) of atmospheric fluorescence light on clear moonless nights. Four fluorescence detector sites surround the surface detector, which has an enclosed area of 3~000~km$^{2}$.

Observations began in 2004 with a partial array and the gradual deployment of the designed Observatory was completed in April 2008 with 1~600 surface detectors and 24 fluorescence telescopes. After additional enhancements to extend the energy coverage to $\sim 10^{17}$~eV, more than 1~660 water--Cherenkov stations and 27 fluorescence telescopes are installed. The Pierre Auger Collaboration is a community of more than 500 members from 19 countries.

Besides its huge aperture (volume of atmosphere and solid angle available for shower detection), the uniqueness of the Pierre Auger Observatory lies in the simultaneous observation of cosmic ray showers with two complementary techniques. Reconstructions combining surface and fluorescence data provide an opportunity to calibrate the surface detector (SD) signal using a shower energy which is measured very precisely by the fluorescence detector (FD). Moreover, in the, so--called, hybrid mode, timing information about the shower arrival on the ground from at least one SD station is used as additional information reconstructing the shower longitudinal profile with the FD. The FD provides a very precise measurement of the properties of the shower electromagnetic component, which contains $\sim$90\% of its energy. 

The processes of managing and reconstructing the shower properties from the measured data are described in the following section. The third section includes specification of the Monte Carlo simulations of UHECR showers. Selected results obtained using the measured and simulated data of the Pierre Auger Observatory are mentioned in Section~4. The conclusions are presented in Section~5.

\section{Measured Data}
Each water--Cherenkov station samples signals from three photomultipliers (PMTs) with a frequency of 40~MHz. The signal is induced by both electromagnetic particles and muons. The time stamp is obtained from GPS. When one of the local trigger conditions on the PMT signals is fulfilled, the signal traces of three PMTs are sent via the communication antenna to the communication towers placed near the fluorescence detector buildings and finally to the central data acquisition system (CDAS), when further physics triggers are tested. Altogether, the SD raw data has to pass a five--level hierarchical trigger system \cite{SdTrigger} to be stored in data centers.

\begin{figure}[!ht]
\centering
\includegraphics[width=0.7\textwidth]{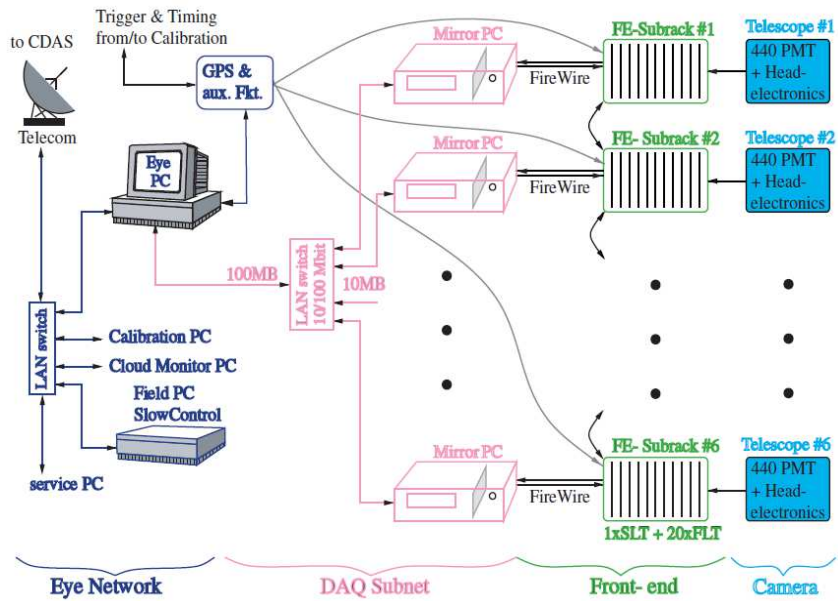}
\caption{Schematic description of readout data flow at one fluorescence detector site. Picture comes from \cite{FdPaper}.}
\label{FdDAQ}
\end{figure}

Each FD telescope collects fluorescence light in a camera consisting of 440~PMTs that are installed in the focal plane of the telescope. Signal traces from PMTs along a candidate shower track are also sent to the CDAS as illustrated in Fig.~\ref{FdDAQ}. FD raw data has to satisfy a three--level hierarchical trigger system \cite{FdPaper} to be stored in data centers. The FD measurement is accompanied by regular monitoring of the atmosphere \cite{AtmMon}.

The SD and FD raw data from CDAS are copied to two facilities, in Lyon and Fermilab, via high speed Internet data--link. The overall size of SD and FD recorded raw data is a few TB per year. One typical shower has size $\sim$hundreds of~kB.

The raw data in .root format are reconstructed using independent software coded in C++: Offline \cite{Offline} for SD and FD and Herald \cite{Herald} for SD. The Offline software is a powerful tool for the simultaneous reconstruction of SD and FD data. It has access to CDAS data and to many formats of simulated showers as well as to databases of calibration constants, atmospheric conditions etc. It has a modular structure that allows a simple implementation of additional reconstruction algorithms or detector enhancements. For the practical purpose of faster and simpler physics analyses, the package Advanced Data Summary Tree (ADST) \cite{Observer} was developed to store all the relevant information including the shower properties reconstructed using the Offline. A typical ADST shower reconstructed in SD and FD has a size of $\sim$10~kB. The ADST data are stored in Karlsruhe on an ftp server with the overall size $\sim$70~GB per one production of all types of data to mid--2014.

\section{Simulated Data}
The simulation of detector signals proceeds in two steps. First, showers of typically $10^{10}$ particles for UHECR energies are simulated with the 3D simulation program CORSIKA \cite{CORSIKA}. It uses different models of hadronic interactions for high ($E_{lab}$ above 80~GeV) and low energy ($E_{lab}$ below 80~GeV) intervals. A so--called thin sampling (thinning) \cite{Thinning} is applied to save computational power. It assigns a higher weight to a randomly selected particle from a sample of secondary particles with total energy higher than an adjustable fraction of the primary energy, while all other secondaries are discarded. The typically used thinning level is $10^{-5}$. A typical shower generated at energy $10^{20}$~eV utilizes approximately 20 hours of CPU time of 1 core on a current processor even with the speed gained from thinning. In the second step, the simulations of detector responses of CORSIKA showers are performed using corresponding modules in the Offline. The required CPU time is smaller for lower energies but rises steeply and reaches values comparable to the CORSIKA part for energies $10^{20}$~eV. 

Because there are several models to compare and many parameters to tune, we need a significant computing and storage capacity for these simulations. Local clusters available to some members of the collaboration were used. Each cluster might have different local setup and operating environment; validation of results was difficult. The unification came with the computing grid solution inspired by LHC experiments. The Prague group created a grid Virtual Organization (VO) \emph{auger} in 2006 and started to run a common production of libraries of simulated showers using a custom production system based on shell scripts \cite{schovan2010}. Later the group from the University of Granada took the responsibility for the official production and developed a second version of the production system with MySQL database back--end \cite{julio2012}. The VO \emph{auger} became one of the biggest users of the EGI grid resources after the LHC VOs \cite{EGIacc}. The success of grid usage was possible thanks to a similar workflow to high energy physics and thanks to resources from many centers where collaboration members were connected to. A significant fraction of CPU time was also provided by non--pledged resources. The stored output of simulated showers on the GRID is currently almost 500~TB in 36 million files.

The success in getting distributed resources brought problems with their management. It was shown that pilot jobs lead to better efficiency and higher usage. Also basic tools provided by the grid middleware for data management are not sufficient for large collections of the size of the VO \emph{auger} data. The DIRAC interware \cite{DIRAC} was tested first as a replacement for the file catalog and later also for the production system. We plan to do a complete migration in 2015.

\section{Physics with Measured and Simulated Data}
Following features of UHECR showers were observed with the Pierre Auger Observatory data and their comparison with MC predictions so far:
\begin{itemize}
\item \textbf{Spectrum:} The so called ankle at energy $10^{18.4}$~eV and steep suppression beyond $10^{19.6}$~eV are observed with high statistical significance \cite{Spectrum}.
\item \textbf{Mass Composition:} A light composition occurs at energies up to the ankle. A transition towards heavier nuclei with increasing energy seems to be present at energies beyond the ankle \cite{Xmax}.
\item \textbf{p--p cross--section at 57~TeV:} The total p--p cross--section was inferred \cite{CrossSection}. Its value lies approximately in the range that is adopted by the models of hadronic interactions.
\item \textbf{Muon Excess:} The muon excess in measured data indicates a lack of understanding for all models of hadronic interactions \cite{Jeff,Balasz,Inclined}.
\item \textbf{Photons and Neutrinos Searches:} No observation of a photon or neutrino shower has been identified yet. Limits on fluxes of photons exclude Top--down models \cite{PhotonFluxLimit}. 
\item \textbf{Anisotropy:} There is no excess of neutrons from the Galactic plane at EeV energies \cite{NeutronsLimit}. The only anisotropy at the highest energies (above~$\sim$55~EeV) that is stable in time is observed in the 20$^{\circ}$ vicinity of the closest AGN: Cen~A \cite{CenA}. The level of correlation with known nearby AGNs from the VCV catalog decreased with time \cite{CenA} and does not increase with energy \cite{AnisotropyICRC13}.
\end{itemize}

An upgrade of the Pierre Auger Observatory with muon detectors is planned to be built in the future. Reliable determination of the mass composition, especially at the highest energies, with large enough statistics seems to be the crucial point in further studies of UHECR open questions (origins of the spectrum features, UHECR sources, lack of muons in MC simulations). Measurement of the muon component with a large surface array (full duty cycle) could be one way to answer these questions. However, the observed inconsistency of model predictions and measured data for the muon component is an issue that needs to be fixed simultaneously. Accompanying measurements with more types of muon detectors combined with the measurement of the electromagnetic component could be not only beneficial, but also necessary.

\section{Conclusions}
Processing of measured as well as simulated data at the Pierre Auger Observatory is well established. Important findings about cosmic rays of ultra--high energy and hadronic interactions at energies beyond accelerator capabilities came from the unprecedentedly large statistics collected by the observatory during ten years of operation. An upgrade of the observatory with muon detectors could bring more light into the presented results. The modular structure of the Offline software provides an easy implementation of such further detector enhancements. The difficulties to perform large productions of MC libraries of cosmic--ray showers for the Pierre Auger Observatory on local computer centers has led to the unprecedented usage of the GRID infrastructures in the field of astroparticle physics. 
\section*{Acknowledgment}
This work is funded by the Czech Science Foundation grant 14-17501S.

\end{document}